\documentclass[%
reprint,aps,amsmath,amssymb]{revtex4-2}

\usepackage{amsmath}
\usepackage{amssymb}

\usepackage{color}
\usepackage{graphicx}
\usepackage{dcolumn}
\usepackage{bm}
\usepackage[colorlinks=true,citecolor=blue,urlcolor=blue,linkcolor=blue,hyperindex]{hyperref}
\usepackage{subfigure}

\usepackage{mathtools}
\usepackage{float}

\usepackage{mathptmx}
\fontsize{12pt}{14pt}\selectfont

\makeatother

\begin{document}

\preprint{APS/123-QED}

\title{Electric Field Induced Superconductivity in Bilayer Octagraphene}

\author{Yitong Yao$^{1}$}
\author{Jun Li$^{2}$}
\author{Jiacheng Ye$^{1}$}
\author{Fan Yang$^{3}$}
\thanks{}
\email{Correspondence: yangfan\_blg@bit.edu.cn}
\author{Dao-Xin Yao$^{1}$}
\thanks{}
\email{Correspondence: yaodaox@mail.sysu.edu.cn}
\affiliation{
$^{1}$Center for Neutron Science and Technology, Guangdong Provincial Key Laboratory of Magnetoelectric Physics and Devices, State Key Laboratory of Optoelectronic Materials and Technologies, School of Physics, Sun Yat-Sen University, Guangzhou 510275, Peoples Republic of China\\
$^{2}$Key Laboratory for Microstructural Material Physics of Hebei Province, School of Science, Yanshan University, Qinhuangdao 066004, China.\\
$^{3}$School of Physics, Beijing Institute of Technology, Beijing 100081, China}

\begin{abstract}
We investigate the energy bands, magnetism, and superconductivity of bilayer octagraphene with A-A stacking under a perpendicular electric field.  A tight-binding model is used to analyze the band structure of the system. The doubling of the unit cell results in each band of the single layer splitting into two. We find that applying a perpendicular electric field increases the band splitting. As the electric field strength increases, the nesting of the Fermi Surface(FS) weakens, eventually disrupting the antiferromagnetic order and bilayer octagraphene exhibits superconductivity. Spin fluctuations can induce unconventional superconductivity with $s^\pm$-wave pairing. Applying a perpendicular electric field to bilayer octagraphene parent weakens the nesting of the FS, ultimately killing the spin-density-wave (SDW) ordered state and transitioning it into the superconducting state, whichworks as a doping effect. We use the random-phase approximation approach to obtain the pairing eigenvalues and pairing symmetries of the perpendicular electric field-tuned bilayer octagraphene in the weak coupling limit. By tuning the strength of the perpendicular electric field, the critical interaction strength for SDW order can be modified, which in turn may promote the emergence of unconventional superconductivity.

\end{abstract}

\maketitle

\section{\label{sec:intro}Introuction}

Superconductivity has long been a prominent area of research in physics. Among the various branches, two-dimensional (2D) superconductivity has emerged as a frontier field, attracting significant attention. Experimental techniques such as molecular beam epitaxy \cite{1,2}, mechanical exfoliation \cite{3,4}, and field-effect transistors \cite{5,6} have been successively applied to the research of 2D superconductivity. For example, in the layered superconductor \textup{NbSe$_2$}, the superconducting critical temperature (T$_C$) decreases as the material thickness is reduced \cite{7}. Additionally, superconductivity has been observed at the interface between \textup{LaAlO$_3$} and \textup{SrTiO$_3$} \cite{8}. Although these materials are not inherently 2D, the observed phenomena at the interface are indicative of 2D superconductivity \cite{11}.

Research on 2D superconductivity has evolved beyond the mere investigation of the effect of material thickness on superconductivity. Recent studies have expanded to include methods for tuning 2D superconductors through pressure \cite{12,pan2015pressure,ao2023valley,kang2015superconductivity,qi2016superconductivity,shao2024possible}, doping \cite{13,li2024tunable}, and external fields \cite{11,shao2024possible,14,15,16,cho2016nanoscale,liu2022superior,tsen2016nature,stojchevska2014ultrafast}. For instance, a superconducting gap of 0.9 meV has been observed in graphene doped with Li \cite{13}. When modulated by electric fields and in contact with superconductors, the resistivity of graphene can be reduced to zero \cite{17}. In bilayer graphene, superconductivity can be induced through the magic angle twist of the two layers \cite{18} and twisted trilayer graphene moir{\'e} superlattices \cite{19}. Electric field modulation has proven particularly effective in controlling superconductivity in bilayer twisted graphene \cite{park2021tunable,hao2021electric,dutta2024electric}, offering a powerful tool for further investigation into the underlying superconducting mechanisms.

2D carbon-based materials show great potential for achieving superconductivity. In addition to the common hexagonal $C_6$ ring structures found in graphene, there are several other two-dimensional materials featuring alternative carbon ring structures that exhibit relatively lower stability, including pentagonal $C_5$ and heptagonal $C_7$ rings \cite{crespi1996prediction,deza2000pentaheptite}, as well as square $C_4$ and octagonal $C_8$ rings \cite{bucknum2008squarographites,liu2012structural}. Octagraphene, composed of pure square $C_4$ and $C_8$ rings, has garnered attention due to its unique structure, which has led to its consideration as a novel carbon-based material with potential superconducting properties \cite{liu2012structural,de2019topological}. Furthermore, the synthesis of biphenylene, a two-dimensional carbon allotrope with a 4-6-8 ring structure, represents a significant breakthrough in the synthesis of octagraphene-like materials \cite{fan2021biphenylene}. Notably, one-dimensional carbon nanoribbons with four-membered and eight-membered rings have also been successfully synthesized \cite{37}.

In our previous work on single-layer octagraphene \cite{20}, we determined that the undoped material adopts an antiferromagnetic (AFM) state, as revealed by random phase approximation (RPA) calculations, with a wave vector $\mathbf{Q} = (\pi, \pi)$. This state exhibits perfect Fermi surface (FS) nesting, which is significant because it suggests behavior akin to cuprate and iron-based superconductors \cite{21,22}, which are known for their potential to host high-temperature superconductivity. To further modulate the system, we applied electron doping, which destroyed the AFM ordered phase and induced spin fluctuations, ultimately leading to high-temperature superconductivity with $s^{\pm}$-wave pairing symmetry. However, in practise, the chemical doping without inducing disorders cannot be easily realized, which drives us to find substitutive approaches to realize the superconductivity in the system.

In this paper, we investigate the energy bands, magnetism, and superconductivity of bilayer octagraphene with A-A stacking under a perpendicular electric field. Unlike doping, the application of a perpendicular electric field results in an increase in energy band splitting rather than a displacement of the entire energy band. As the electric field strength increases, the bandwidth broadens, the nesting of the FS weakens, and the magnetic properties diminish. However, both methods ultimately disrupt the AFM order of the system, leading to spin fluctuations that induce an $s^{\pm}$-wave paring unconventional superconductivity. Using the RPA approach to model the effects of a perpendicular electric field on bilayer octagraphene, we find the largest pairing eigenvalue $\lambda \approx 0.32$. Our study demonstrates that the perpendicular electric field can serve as a viable approach to induce unconventional superconductivity in the bilayer octagraphene.

\begin{figure}[t]
 
	\centering
\subfigure[]{\label{fig1a}\includegraphics[width=0.24\textwidth]{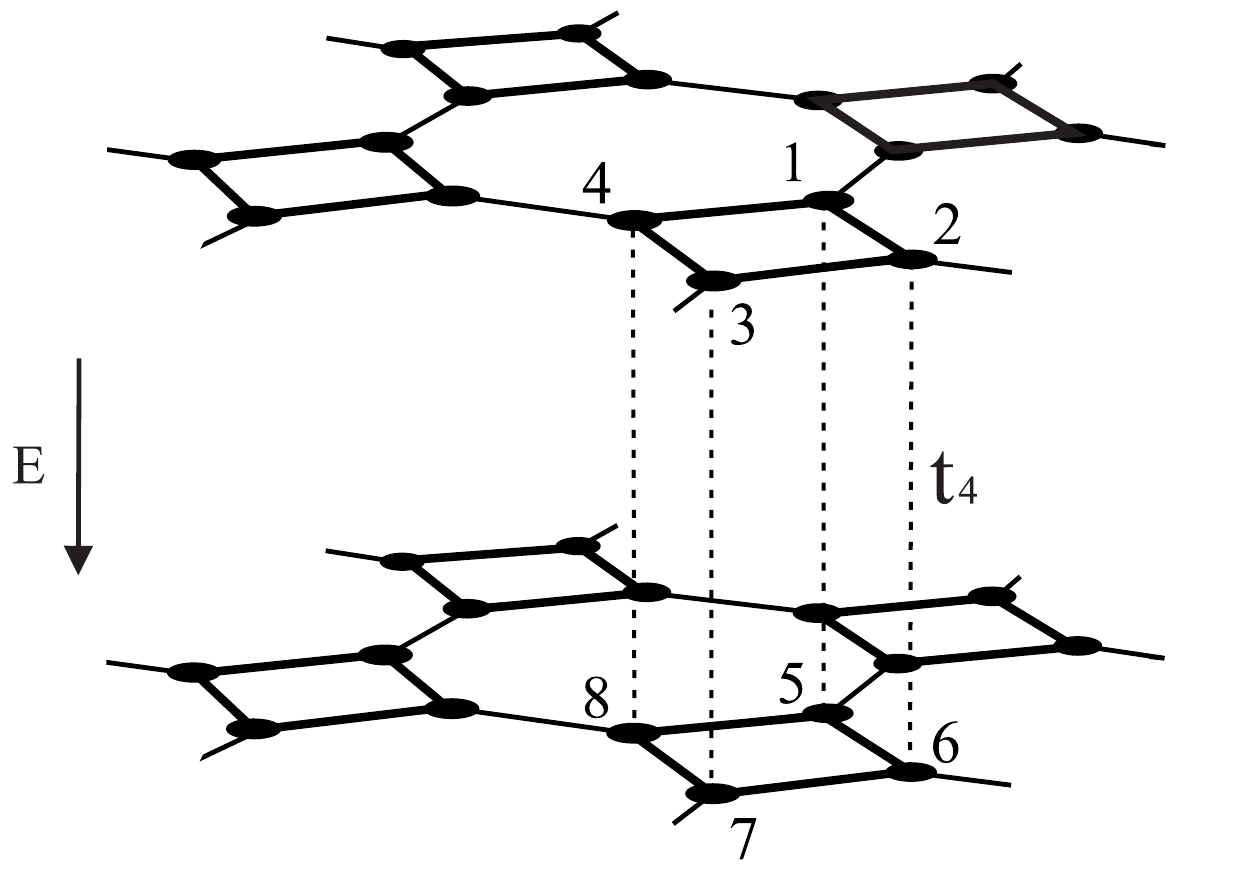}}\subfigure[]{ \label{fig1b}\includegraphics[width=0.24\textwidth]{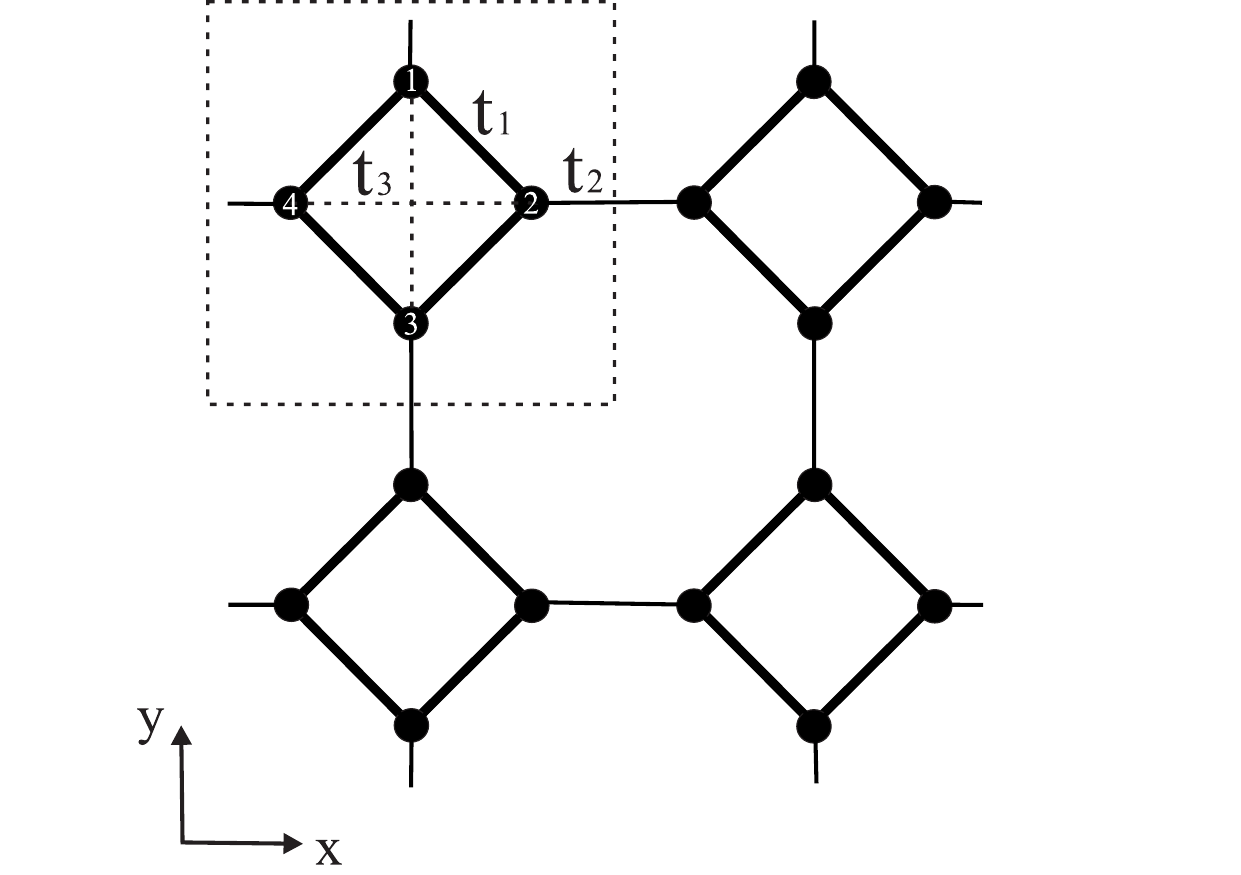}}\\
\subfigure[]{\label{fig1c}\includegraphics[width=0.24\textwidth]{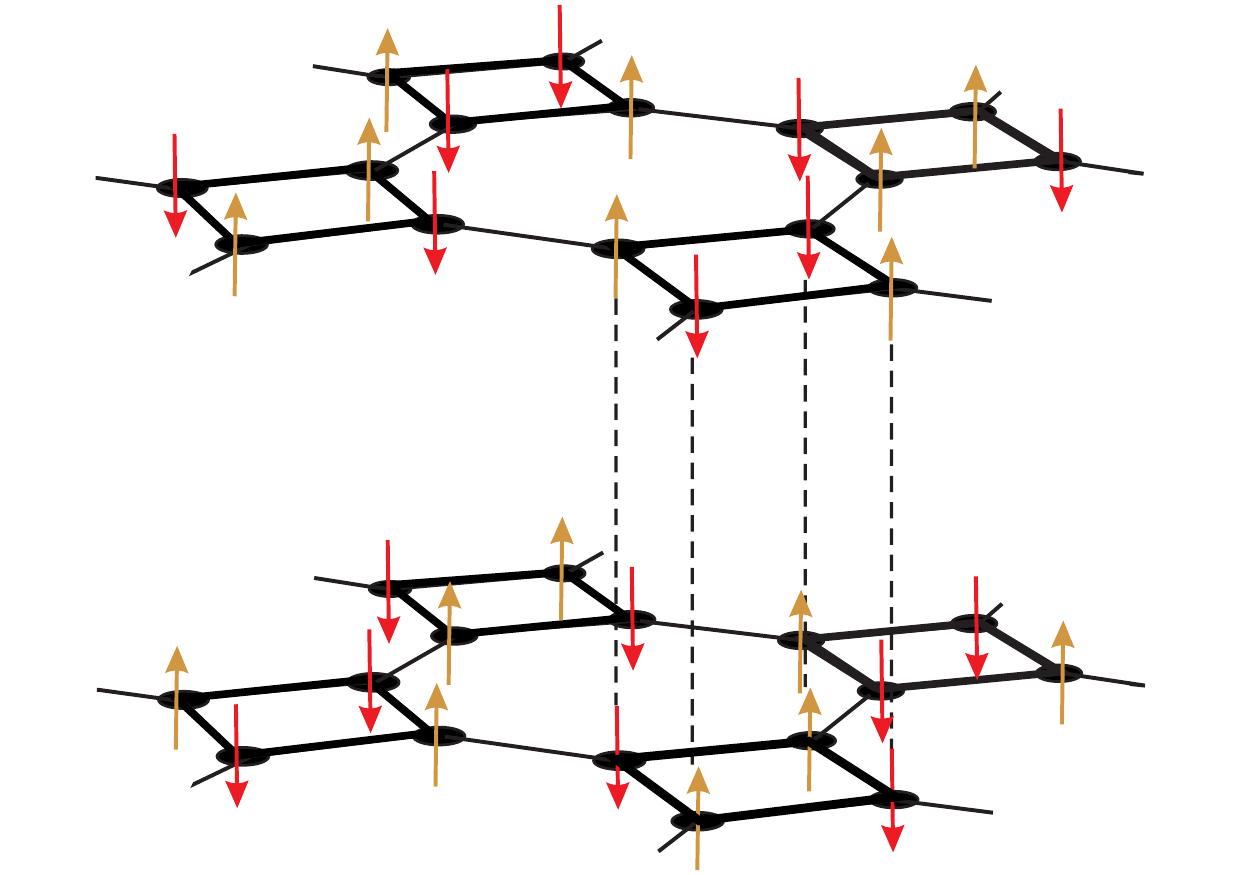}}
	\caption{ (a) The structure of bilayer octagraphene. The arrow dnotes the direction in which the perpendicular electric field is being applied. (b) Top view of bilayer octagraphene, $t_1$,  $t_2$ , $t_3$ and $t_4$ correspond to the intra-square, inter-square, intra-square diagonal position and inter-layer energies, respectively. (c)  The antiferromagnetic ordered spin pattern in the bilayer octagraphene without electric field. }
      \label{Fig:1}
\end{figure}

\section{\label{s}The Model and Theory}

\subsection{\label{sec:model}The Model}

We build a tight-binding (TB) model to describe bilayer octagraphene. As shown in Fig.\ref{Fig:1}, single-layer octagraphene is composed of squares and regular octagons, as for the bilayer structure contains 8 carbon atoms in a unit cell, including 4 carbon atoms in each layer. Consider hopping integrals $t_1$ -- $t_4$ in this model, $t_4$ represents the nearest neighbor (NN) hopping integrals between two layers, while $t_{1}$ -- $t_3$ are the intra-layer hopping integrals, corresponding to intra-square NN hopping integrals, inter-square NN hopping integrals and intra-square next nearest neighbor (NNN) hopping integrals, respectively. Noting that, due to the $C_{4v}$ symmetry of the octagraphene, there are three probable stacking modes between two layers. The most probable stacking mode is A-A stacking, which is defined as (0, 0) relative shifts between two layers, as it is stabler than A-B (0.5, 0.5) stacking and A-C (0, 0.5) stacking in our previous calculations~\cite{23}. Therefore, the $t_4$ is perpendicular to the layers. Each carbon atom of octagraphene contributes an electron in one 2$p_z$ orbital. We consider the system at half-filling. The TB Hamiltonian of the bilayer octagraphene can be expressed as

\begin{equation}
	\label{eq:tb}
     \begin{split}
	H_{TB} =-\sum_{i, j, \sigma,\alpha} t_{ij} c_{i \sigma \alpha}^{\dagger} c_{j \sigma \alpha} -\sum_{i, \sigma} t_4 c_{i \sigma 1}^{\dagger} c_{i \sigma 2} \\
    + \frac{1}{2} V \sum_{i \sigma} n_{i \sigma 1} - \frac{1}{2} V \sum_{i \sigma} n_{i \sigma 2} +H.c.,
     \end{split}
\end{equation}

\noindent where $t_{ij}$ represents intralayer hopping integrals $t_1$ -- $t_3$, $t_4$ represents interlayer hopping integral, $\alpha$ = 1, 2 for top and bottom layers, $c_{i \sigma}^{\dagger}$ ($c_{i \sigma}$) creates (annihilates) an electron at site $i$ in the layer $\alpha$ with spin $\sigma$. $V$ represents the potential energy difference between the two layers, with the plus sign for the top layer and the minus sign for the bottom layer. Because there are eight carbon atoms in a unit cell, the TB Hamiltonian can be written as an eight-order matrix
\begin{equation}
    \centering
    \label{eq:hamiltonian}
    \widetilde{H}_{TB}=\left[\begin{array}{cc} A_1 & B \\ B & A_2 \end{array}\right],
\end{equation}
\noindent where $A_1$, $A_2$ and $B$ are written as
\small{
\begin{equation}
\setlength{\arraycolsep}{1.0pt}
    \centering
    \label{eq:A1}
    A_1 = -\left[\begin{array}{cccc} {-\frac{1}{2} V} & {t_1} & {t_2 e^{i k_{y}} +t_3} & {t_1} \\  {t_1} & {-\frac{1}{2} V} & {t_1} & {t_2 e^{i k_{x}} +t_3} \\ {t_2 e^{-i k_{y}} +t_3} & {t_1} & {-\frac{1}{2} V} & {t_1} \\  {t_1} & {t_2 e^{-i k_{x}} +t_3} & {t_1} & {-\frac{1}{2} V} \end{array}\right],
\end{equation}}

\small{
\begin{equation}
\setlength{\arraycolsep}{1.0pt}
    \centering
    A_2 = -\left[\begin{array}{cccc} {\frac{1}{2} V} & {t_1} & {t_2 e^{i k_{y}} +t_3} & {t_1} \\ {t_1} & {\frac{1}{2} V} & {t_1} & {t_2 e^{i k_{x}} +t_3} \\ {t_2 e^{-i k_{y}} +t_3} & {t_1} & {\frac{1}{2} V} & {t_1} \\  {t_1} & {t_2 e^{-i k_{x}} +t_3} & {t_1} & {\frac{1}{2} V} \end{array}\right],
\end{equation}}

\begin{equation}
    \centering
    B =- \left[\begin{array}{cccc} {t_4} & {0}  & {0}  & {0} \\  {0} & {t_4}  & {0}  & {0} \\ {0}  & {0} & {t_4} & {0} \\ {0}  & {0} & {0} & {t_4} \end{array}\right].
\end{equation}

\begin{figure}[t]

	\centering

  \subfigure[]{\label{fig2a}\includegraphics[width=0.24\textwidth]{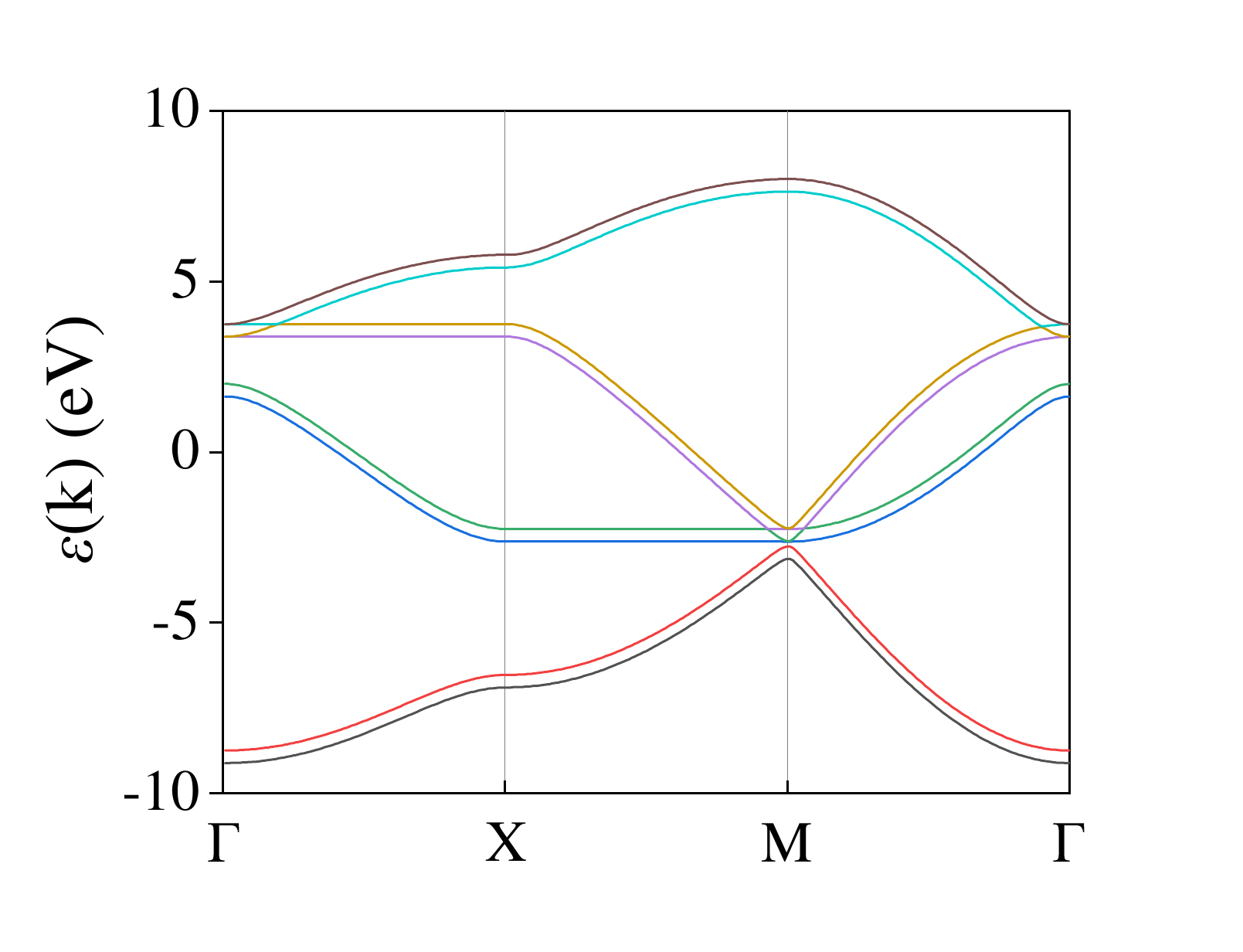}}\subfigure[]{ \label{fig2b}\includegraphics[width=0.24\textwidth]{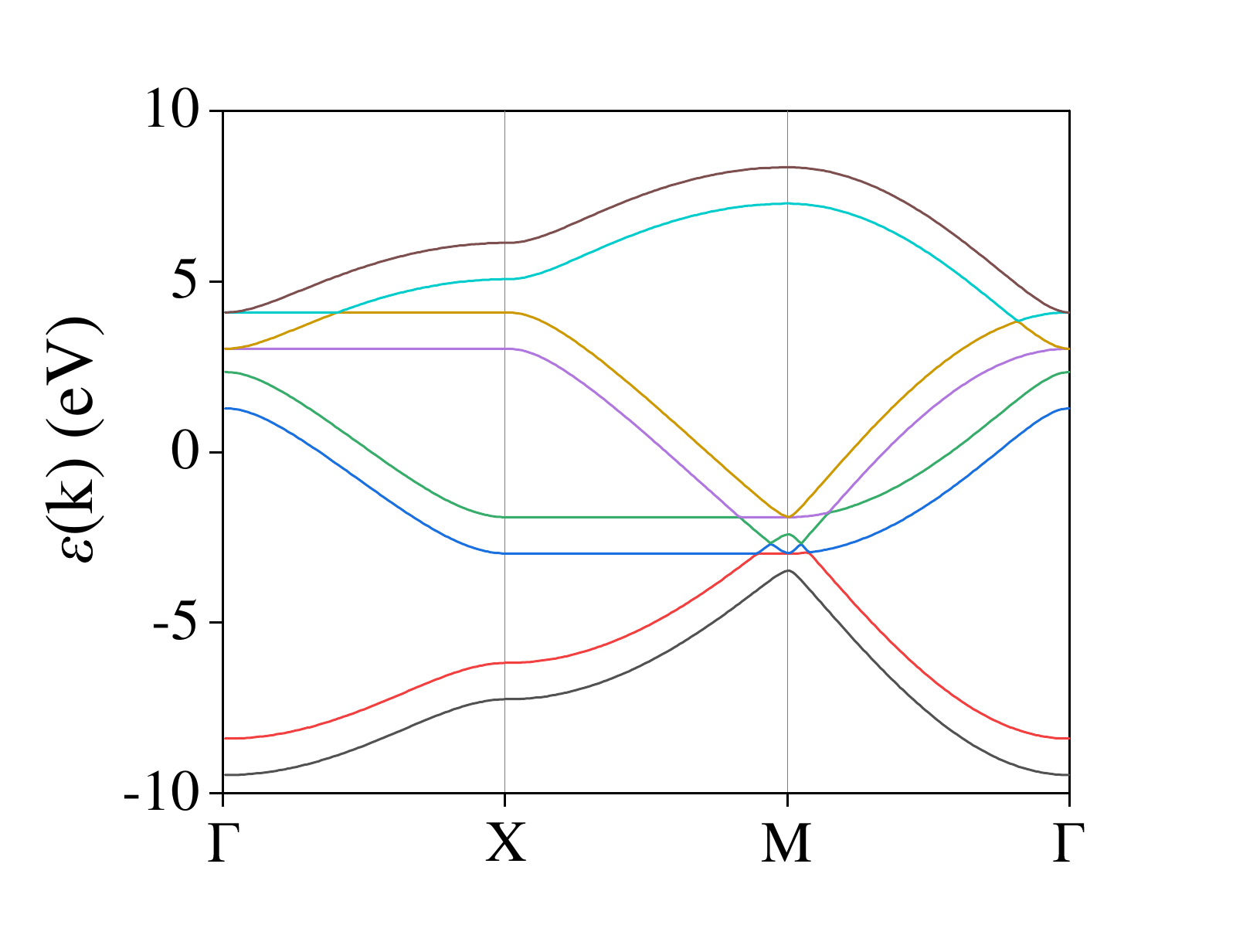}}\\
 \subfigure[]{\label{fig2c}\includegraphics[width=0.22\textwidth]{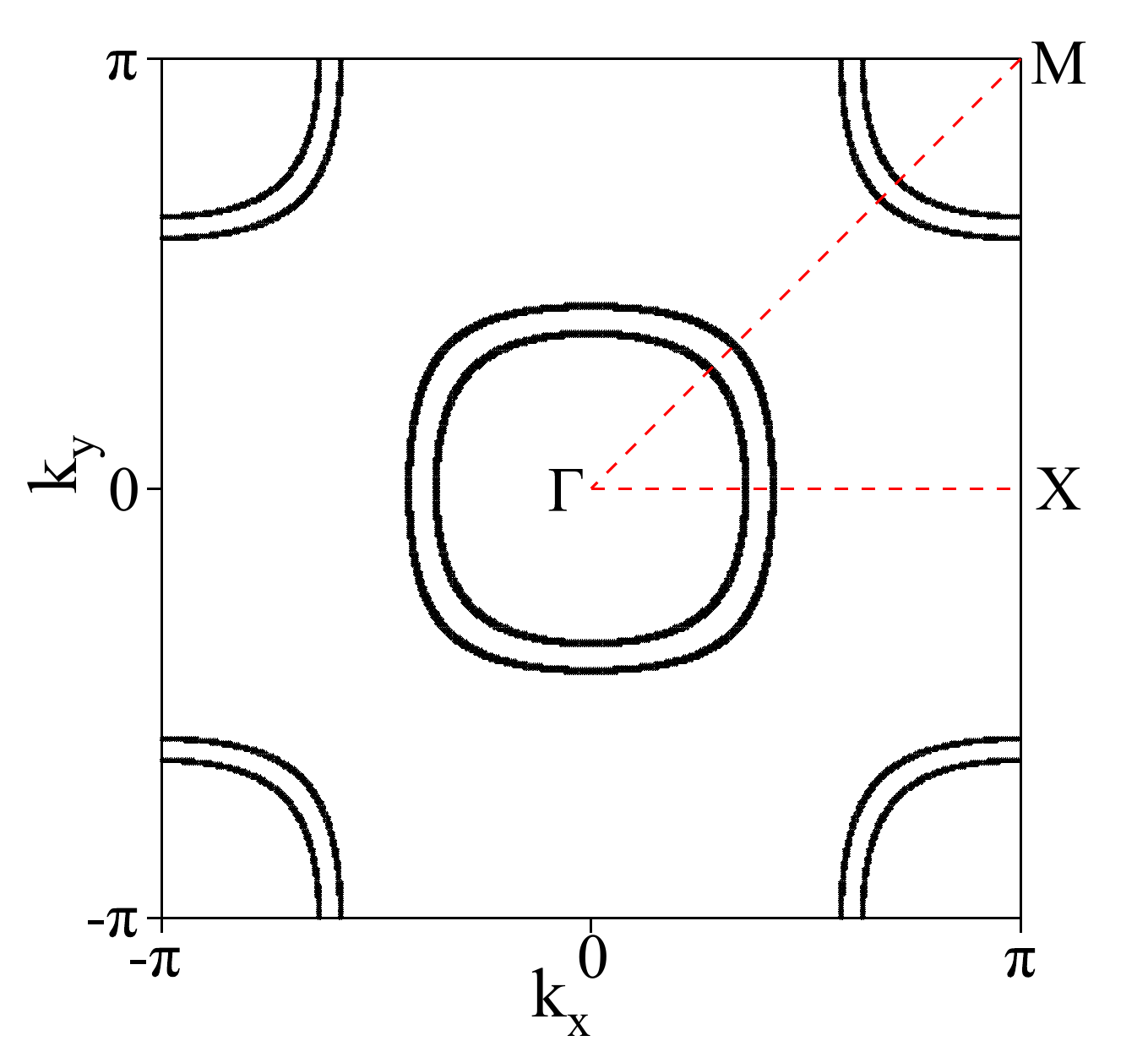}} ~~~~\subfigure[]{\label{fig2d}\includegraphics[width=0.22\textwidth]{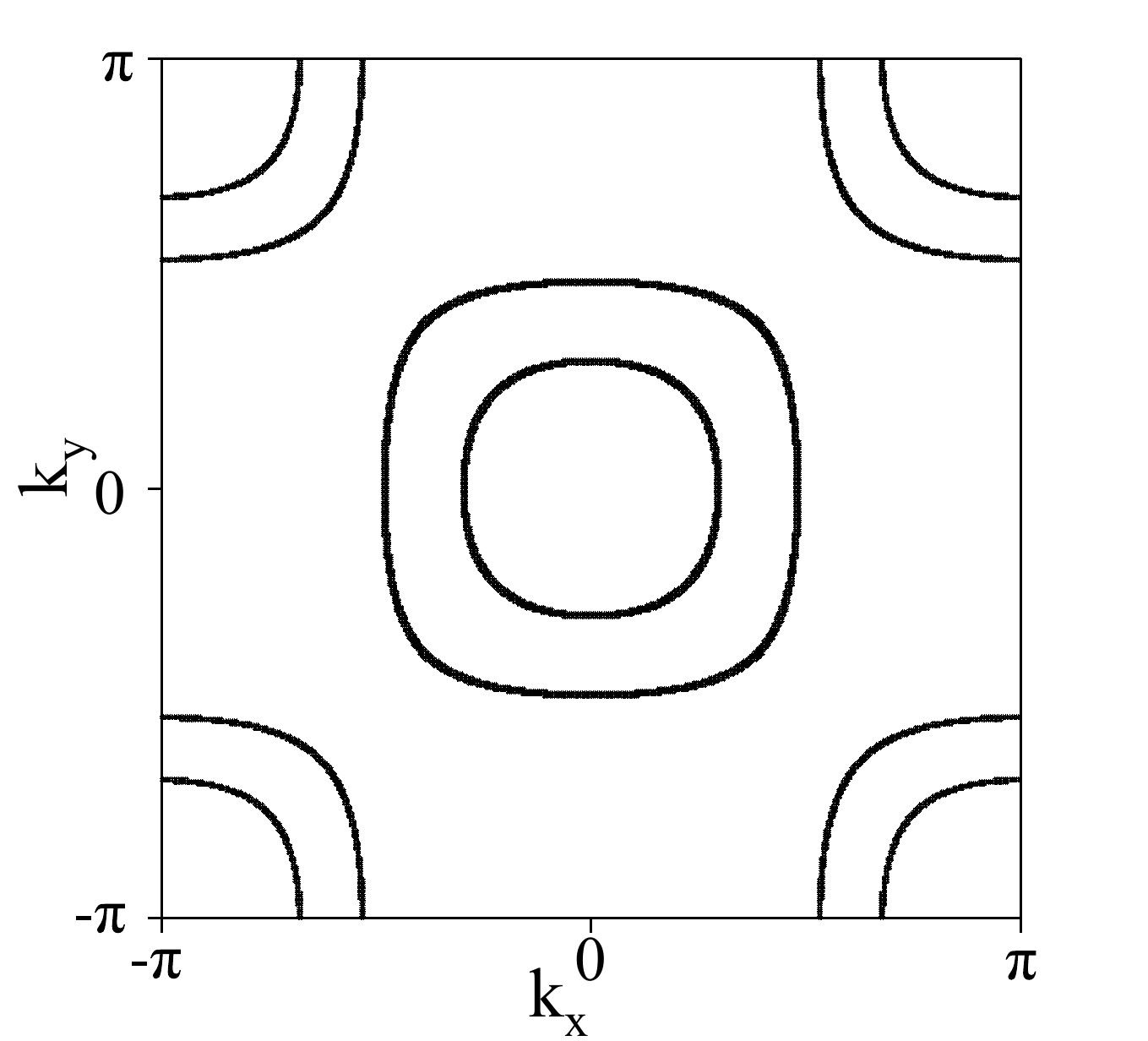}}
	
	\caption{(a) and (b) Band structures of bilayer octagraphene at $V$ = 0 and $V$ = 1.0 eV. (c) and (d) FS of bilayer octagraphene at $V$ = 0 and $V$ = 1.0 eV.}
     \label{Fig:2}
\end{figure}

We get the detailed parameters of hopping integral in our previous work by density functional theory calculations: $t_1$ = 2.685 eV, $t_2$ = 3.001 eV, $t_3$ = 0.558 eV, and $t_4$ = 0.184 eV\cite{23}.

\begin{figure}[t]
\centering
	
 \subfigure[]{\label{fig3a}\includegraphics[width=0.23\textwidth]{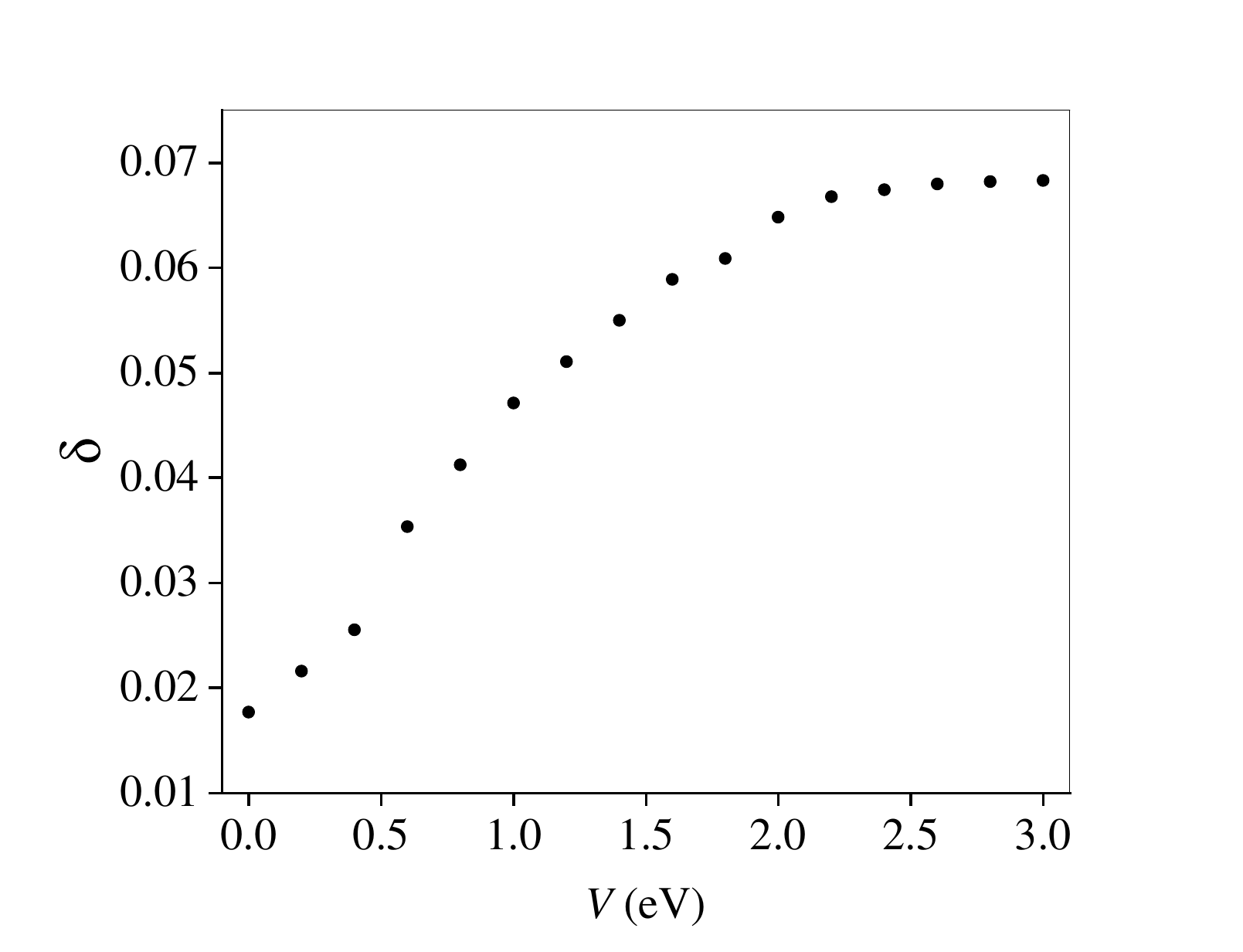}}\subfigure[]{ \label{fig3b}\includegraphics[width=0.23\textwidth]{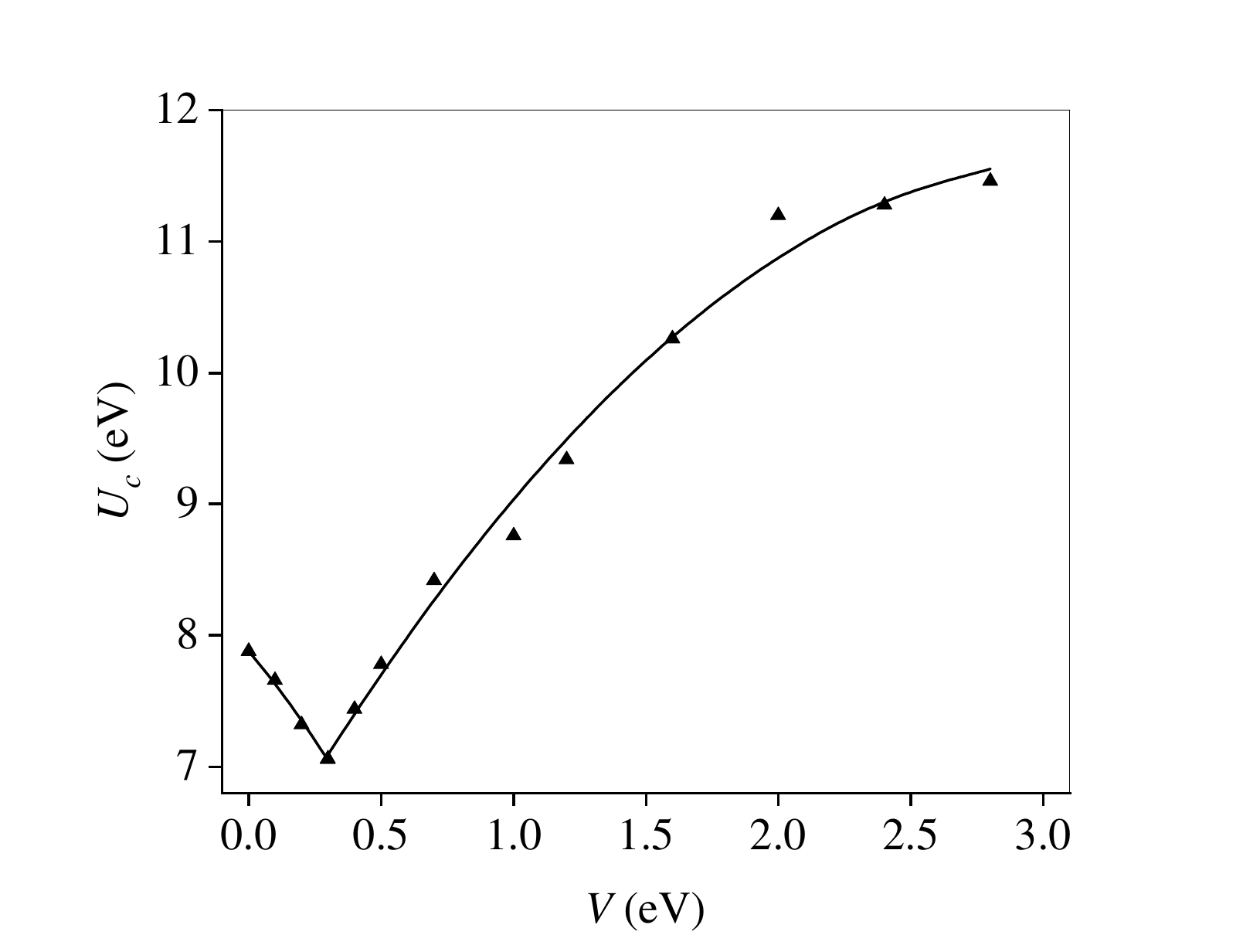}}

  \caption{(a) The deviation of the nesting vector from ($\pi$, $\pi$). (b) RPA-calculated $U_c$ as a function of $V$.}
\label{Fig:3}
  
\end{figure}

Compared with the single layer, each band splits into two.  For the half-filling case, there are two hole pockets around $\Gamma$ point and two nesting pockets around the $M$ point. Fig.\ref{Fig:2} shows the band structure and FS for $V$ = 0 and $V$ = 1.0 eV. The electric field applied between two layers increases the band splitting width. As $V$ increases, the inner Fermi pockets shrink and the outer pockets expand, the nesting of the FS is weakened. In addition, we calculate the number of electrons before and after applying an electric field and confirmed that the number of electrons is conserved. The pockets centered at $(\pi, \pi)$ are electron pockets, and the ones centered at $(0, 0)$ are hole pockets. It loses the perfect nesting of FS when the single layer is turned into a bilayer. We use $\delta$ to express the deviation of the nesting vector from $(\pi, \pi)$, as shown in Fig.\ref{fig3a}.

 Octagraphene is a graphene-based material, the strong Coulomb repulsions between the 2$p_z$ electrons cannot be ignored. We build a Hubbard model to describe it

\begin{equation}
	\label{eq:U}
	H_{Hubbard}=H_{\mathrm{TB}}+U \sum_{i} \hat{n}_{i \uparrow} \hat{n}_{i \downarrow}.
\end{equation}

\noindent The $U$ term represents the on-site Coulomb repulsive interaction.

\subsection{\label{sec:rpa}RPA approach}

We adopt the standard multi-orbital RPA approach \cite{24,25,26,27,28,29,30,31,32} to treat the weak-coupling limit of the model Eq.\ref{eq:U}. We define the free susceptibility for $U = 0$ as

\begin{equation}
\begin{aligned}
	\label{eq:chi}
	\chi_{l_3, l_4}^{(0) l_1, l_2}\left(\mathbf{q}\right)=&\frac{1}{N} \sum_{\mathbf{k}, \alpha, \beta} \xi_{l_4}^{\alpha}(\mathbf{k}) \xi_{l_3}^{\alpha, *}(\mathbf{k}) \xi_{l_2}^{\beta}(\mathbf{k'}) \xi_{l_1}^{\beta, *}(\mathbf{k'}) \\ &\times\frac{n_{F}\left(\varepsilon_{\mathbf{k'}}^{\beta}\right)-n_{F}\left(\varepsilon_{\mathbf{k}}^{\alpha}\right)}{\varepsilon_{\mathbf{k}}^{\alpha}-\varepsilon_{\mathbf{k'}}^{\beta}}.
\end{aligned}
\end{equation}

For $U > 0$, the RPA spin and charge susceptibility can be written as

\begin{equation}
	\label{eq:chis}
	\chi^{(c/s)}(\mathbf{q})=\left[I\pm\chi^{(0)}(\mathbf{q})\widetilde{U}
	\right]^{-1} \chi^{(0)}(\mathbf{q})
\end{equation}

\noindent where $\chi^{(0)}(\mathbf{q})$ and $\widetilde{U}$ are 64$\times$64 matrices, $\widetilde{U}_{l_3 l_4}^{l_1 l_2}=U \delta_{l_{1}=l_{2}=l_{3}=l_{4}}$, and $I$ is the identity matrix. When $U > 0$, spin fluctuations dominate over charge fluctutions. The critical interaction strength $U_c$ is determined by $det[I-\chi^{0}(\mathbf{q})U]=0$. The RPA approach only works for $U < U_c$. For $U > U_c$ the spin susceptibility diverges, which suggests that long range spin-density-wave (SDW) order emerges with the pattern shown in Fig.~\ref{Fig:1}(c).

\begin{figure}[t]

	\centering
 \subfigure[]{\label{fig4a}\includegraphics[width=0.24\textwidth]{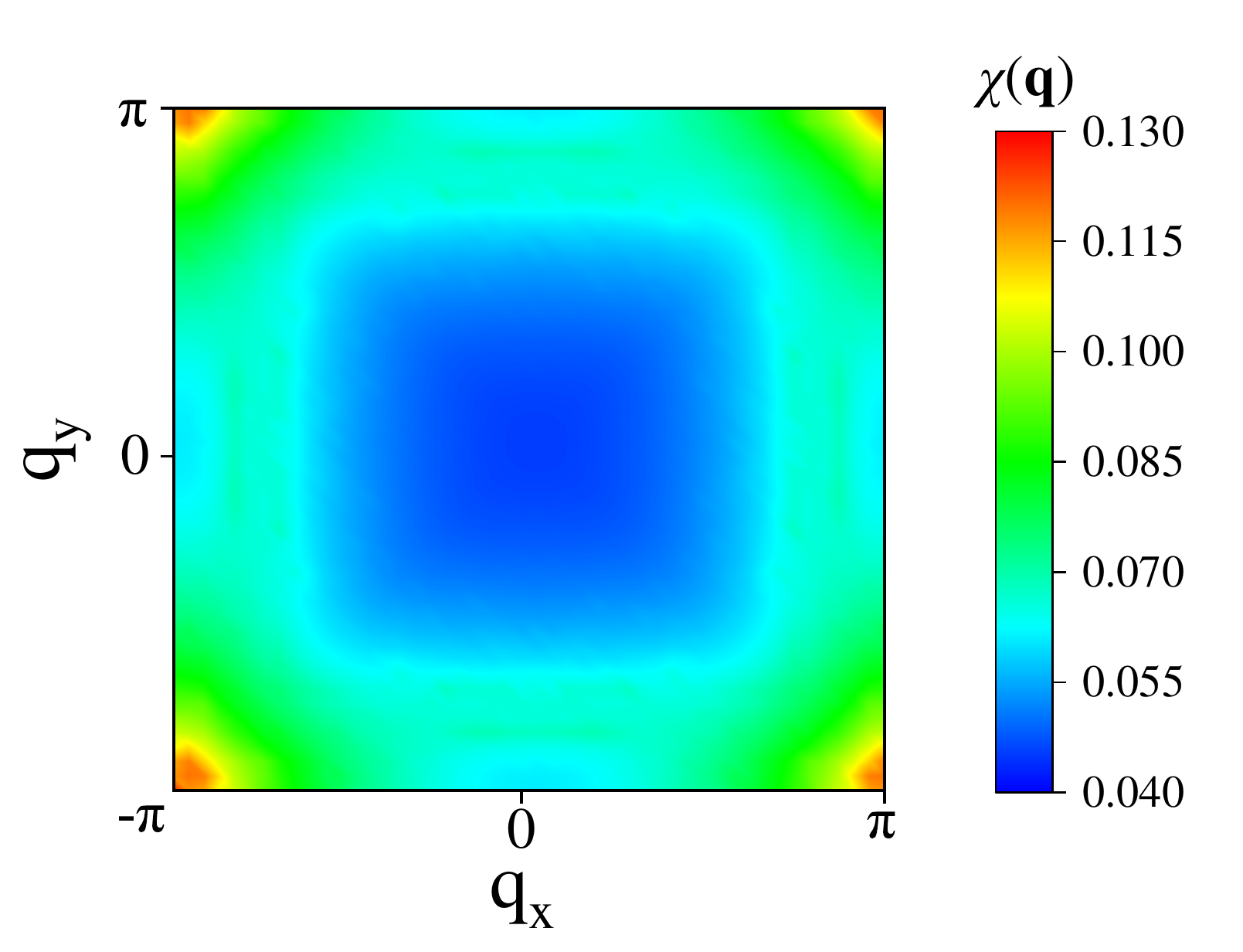}}\subfigure[]{ \label{fig4b}\includegraphics[width=0.24\textwidth]{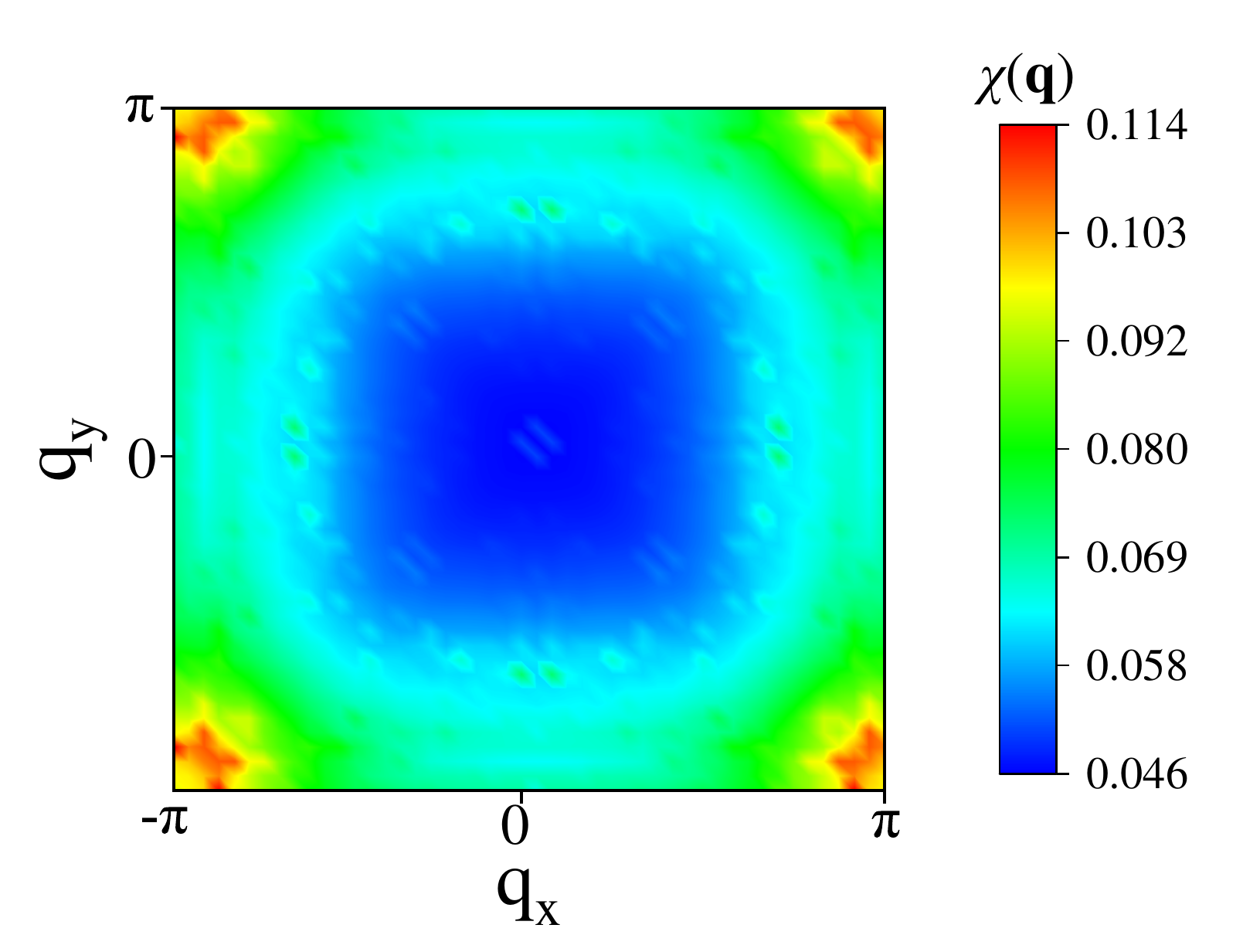}}\\
 \subfigure[]{\label{fig4c}\includegraphics[width=0.24\textwidth]{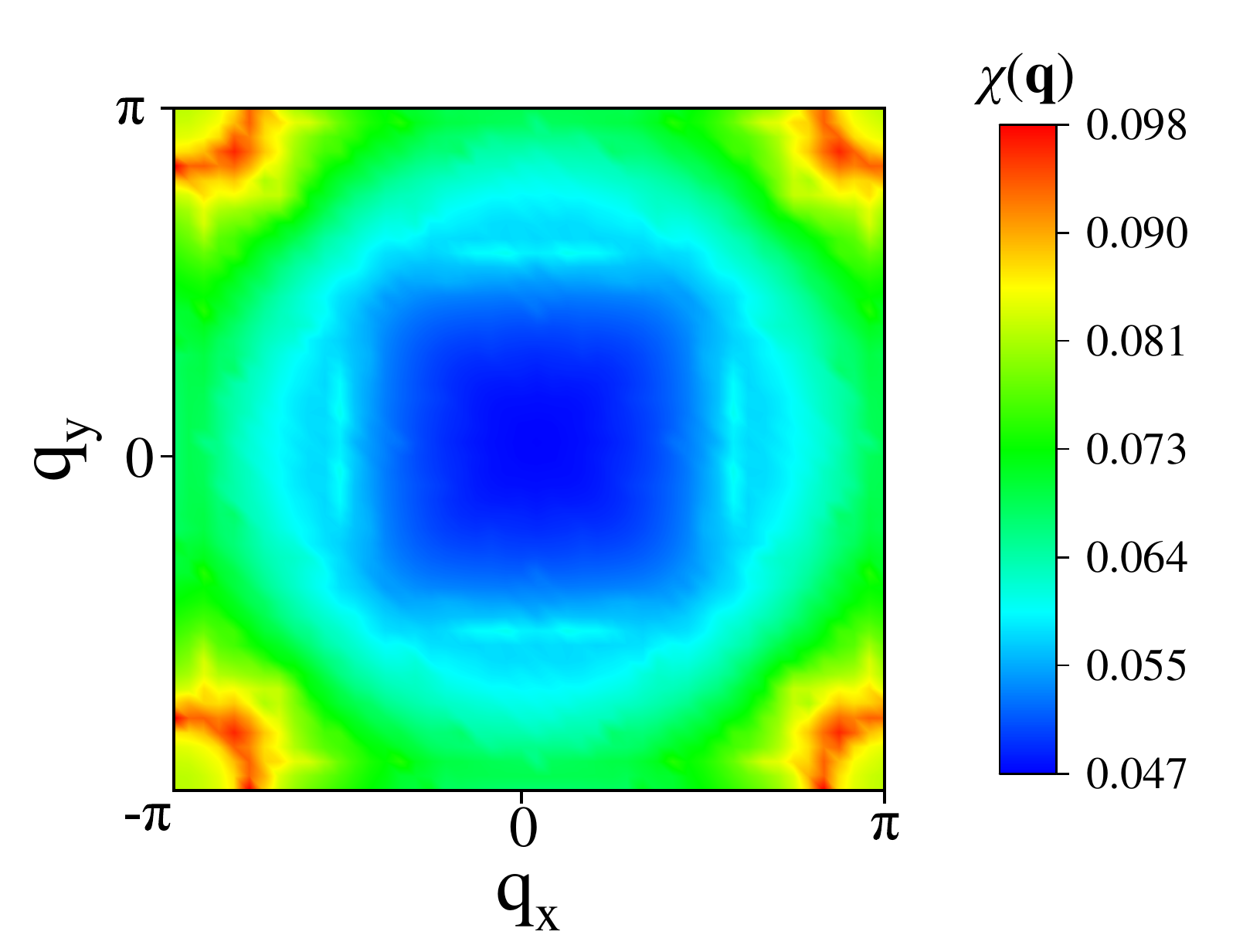}}~~\subfigure[]{\label{fig4d}\includegraphics[width=0.24\textwidth]{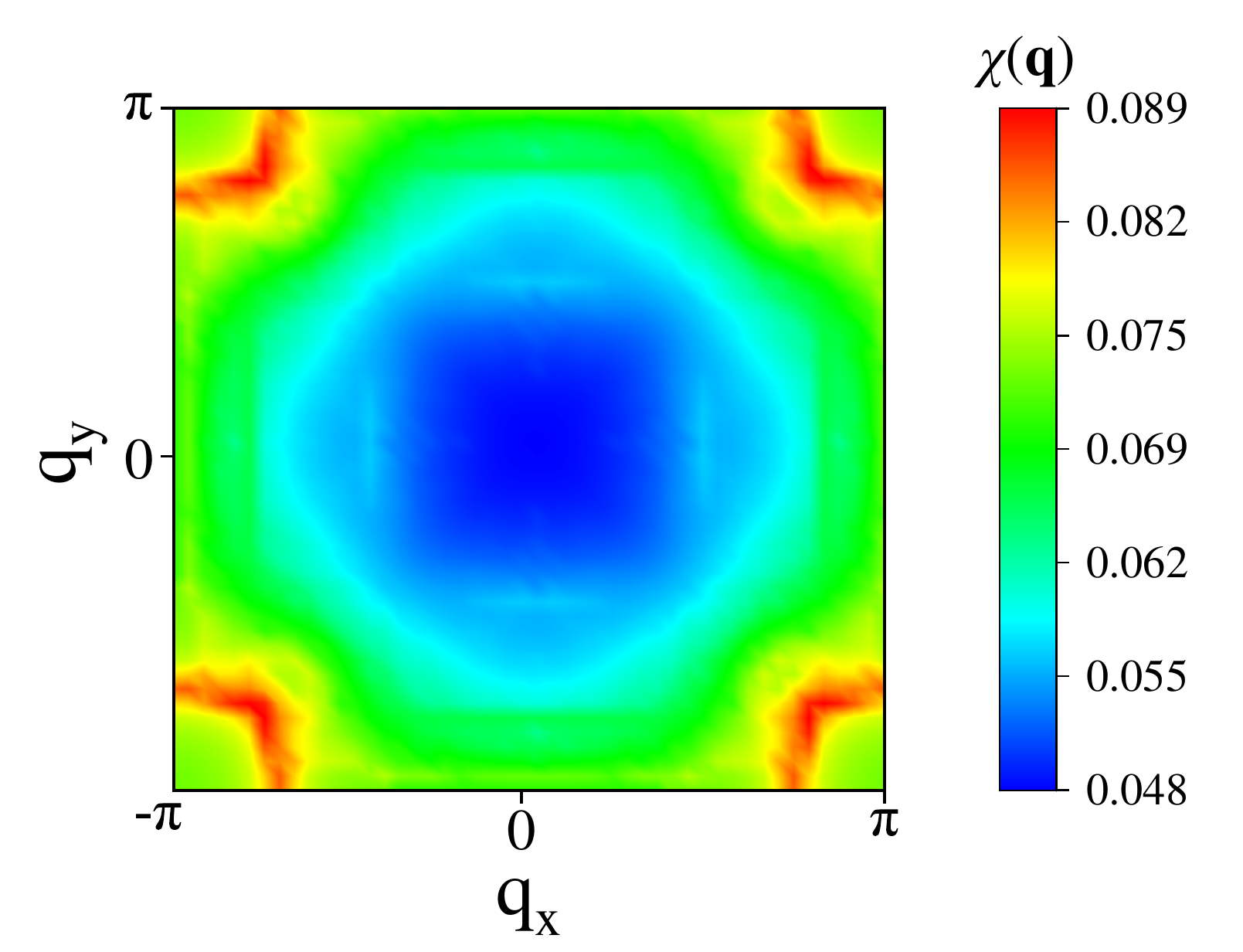}}

	\caption{The $\mathbf{q}$-dependence of the eigen-susceptibilities $\chi$($\mathbf{q}$) in the first Brillouin zone for (a) $V$ = 0, (b) $V$ = 1.0 eV, (c) $V$ = 1.5 eV, (d) $V$ = 2.0 eV, respectively.}
 \label{fig:chi}
\end{figure}

The following linearized gap function is used to determine the $T_c$ and leading pairing symmetry of the system

\begin{equation}
	-\frac{1}{(2 \pi)^{2}} \sum_{\beta} \oint_{F S} d k_{\|}^{\prime} \frac{V^{\alpha \beta}\left(\mathbf{k}, \mathbf{k}^{\prime}\right)}{v_{F}^{\beta}\left(\mathbf{k}^{\prime}\right)} \Delta_{\beta}\left(\mathbf{k}^{\prime}\right)=\lambda \Delta_{\alpha}(\mathbf{k}).
	\label{Eq:gp}
\end{equation}

\noindent where $\alpha$, $\beta$=1,2,3,4 represent the energy bands which cross the Fermi Level, $k_{\|}^{\prime}$ is the component along the FS, and ${v_{F}^{\beta}\left(\mathbf{k}^{\prime}\right)}$ is the fermi velocity at $\mathbf{k}^{\prime}$. We obtain ${V^{\alpha \beta}\left(\mathbf{k}, \mathbf{k}^{\prime}\right)}$ from exchanging spin and charge fluctuations. At the RPA level, the effective interaction for Cooper pairs near FS is

\begin{equation}
	\label{eq:v}
	V_{e f f}=\frac{1}{N}\! \sum_{\alpha \beta, \mathbf{k k}^{\prime}}\! V^{\alpha \beta}\left(\mathbf{k}, \mathbf{k}^{\prime}\right) c_{\alpha}^{\dagger}(\mathbf{k}) c_{\alpha}^{\dagger}(-\mathbf{k}) c_{\beta}\left(-\mathbf{k}^{\prime}\right) c_{\beta}\left(\mathbf{k}^{\prime}\right),
\end{equation}

\noindent which shows a Cooper pair with momentum  $\mathbf{k}$ and orbital $(t, s)$ scattered to $\mathbf{k'}$ and orbital $(p, q)$ by charge or spin fluctuations. The Cooper pair can be spin singlet or triplet. Here,

\begin{equation}
	\begin{aligned}
		&V^{\alpha \beta}\left(\mathbf{k}, \mathbf{k}^{\prime}\right)=
		\\
		&\operatorname{Re} \sum_{p q s t, \mathbf{k} \mathbf{k}^{\prime}} \Gamma_{s t}^{p q}\left(\mathbf{k}, \mathbf{k}^{\prime}, 0\right) \xi_{p}^{\alpha, *}(\mathbf{k}) \xi_{q}^{\alpha, *}(-\mathbf{k}) \xi_{s}^{\beta}\left(-\mathbf{k}^{\prime}\right) \xi_{t}^{\beta}\left(\mathbf{k}^{\prime}\right).
	\end{aligned}
\end{equation}

For the Cooper pairs in the singlet channel, $\Gamma_{s t}^{p q}\left(k, k^{\prime}\right)$ is

\begin{equation}
	\label{eq:lam}
	\begin{aligned}
		\Gamma_{s t}^{p q}\left(k, k^{\prime}\right)=\widetilde{U}_{q s}^{p t}&+\frac{1}{4}\left\{\widetilde{U}\left[3 \chi^{(s)}\!\left(k\!-\!k^{\prime}\right)-\chi^{(c)}\!\left(k\!-\!k^{\prime}\right)\right]\widetilde{U}\right\}_{q s}^{p t}\\&+
		\frac{1}{4}\left\{\widetilde{U}\left[3 \chi^{(s)}\!\left(k\!+\!k^{\prime}\right)-\chi^{(c)}\!\left(k\!+\!k^{\prime}\right)\right]\widetilde{U}\right\}_{q t}^{p s},
	\end{aligned}
\end{equation}

\noindent In the triplet channel, it is

\begin{equation}
	\label{eq:Tc}
	\begin{aligned}
		\Gamma_{s t}^{p q}\left(k, k^{\prime}\right)=&-\frac{1}{4}\left\{\widetilde{U}\left[\chi^{(s)}\left(k-k^{\prime}\right)+\chi^{(c)}\left(k-k^{\prime}\right)\right]\widetilde{U}\right\}_{q s}^{p t}\\&+
		\frac{1}{4}\left\{\widetilde{U}\left[\chi^{(s)}\left(k+k^{\prime}\right)+\chi^{(c)}\left(k+k^{\prime}\right)\right]\widetilde{U}\right\}_{q t}^{p s}.
	\end{aligned}
\end{equation}

By sloving the linearized gap equation (Eq.\ref{Eq:gp}), we can determine the leading pairing symmetry of the system. The eigenvector $\Delta_{\alpha}(\mathbf{k})$ is the superconducting order parameter, which determines the pairing symmetry, and the largest eigenvalue $\lambda$ is related to $T_{c}$ as follows:

\begin{equation}
	\lambda^{-1}=\ln \left(1.13 \frac{\hbar \omega_{D}}{k_{B} T_{c}}\right),
\end{equation}

\noindent where $\omega_{D}$ is the energy scale of spin fluctuation. 

\section{\label{sec:SCgap} Results and discussion}

Fig.~\ref{fig:chi} illustrates the distribution of the eigen-susceptibility $\chi(\mathbf{q})$ in the first Brillouin zone for varying values of the $V$. In the absence of a perpendicular electric field, the susceptibility exhibits a sharp peak at the wave vector $\mathbf{q} = (\pi, \pi)$, corresponding to the inter-unit-cell Ne\'el AFM order, which is characteristic of a checkerboard spin pattern with alternating spins between adjacent unit cells. This result is consistent with the expected magnetic ordering in a system with such a symmetry, where the inter-unit-cell interaction dominates and stabilizes the AFM ordering.

As the strength of the perpendicular electric field is increased, a noticeable change in the distribution of $\chi(\mathbf{q})$ occurs. The previously sharp peak at $(\pi, \pi)$ begins to split and shift, moving towards $(\pi \pm \delta, \pi \pm \delta)$, where $\delta$ is a small deviation determined by the electric field strength. This splitting indicates that the magnetic ordering is no longer strictly inter-uinit-cell Ne\'el antiferromagnetic order, but becomes an incommensurate one. The new spin configuration, which is influenced by both the electric field and the electron-electron interaction, can be analyzed by examining the eigenvectors of the susceptibility matrix $\chi_{m, m}^{(s) l, l}$, which reveal the detailed spin texture that emerges in the presence of these perturbations.

Moreover, as the $V$ increases, the nesting of the FS becomes progressively weaker. This weakening of FS nesting results in a decrease in the peak value of the eigen-susceptibility $\chi(\mathbf{q})$, indicating that the system’s magnetic response is less pronounced. Consequently, the magnetic ordering transitions from a long-range AFM state to a regime dominated by strong spin fluctuations. These fluctuations signal the destruction of the long-range AFM order, and in this new regime, the system is more likely to exhibit superconducting mediated by spin fluctuations.

This analysis suggests that the introduction of a perpendicular electric field and the variation of $V$ provides powerful tuning knobs for controlling the magnetic and superconducting properties of the system, potentially facilitating the realization of  unconventional superconductivity through the modulation of spin fluctuations and FS nesting.

The bilayer octagraphene system hosts the $C_{4v}$ point group symmetry, and its possible superconducting pairing symmetries can be classified according to the irreducible representations (irreps) of this symmetry group. Specifically, the $A_1$ irrep corresponds to the s-wave pairing, characterized by the $C_{4v}$ symmetric gap function; the $B_1$ irrep corresponds to the $d_{x^2 - y^2}$-wave pairing hosting gap function with nodes along the Brillouin zone diagonal direction; the $B_2$ irrep corresponds to the $d_{xy}$-wave, hosting a gap function with nodes on the $x$- and $y$- axes. 

\begin{figure}[t]
\centering

 \subfigure[]{\label{fig5a}\includegraphics[width=0.23\textwidth]{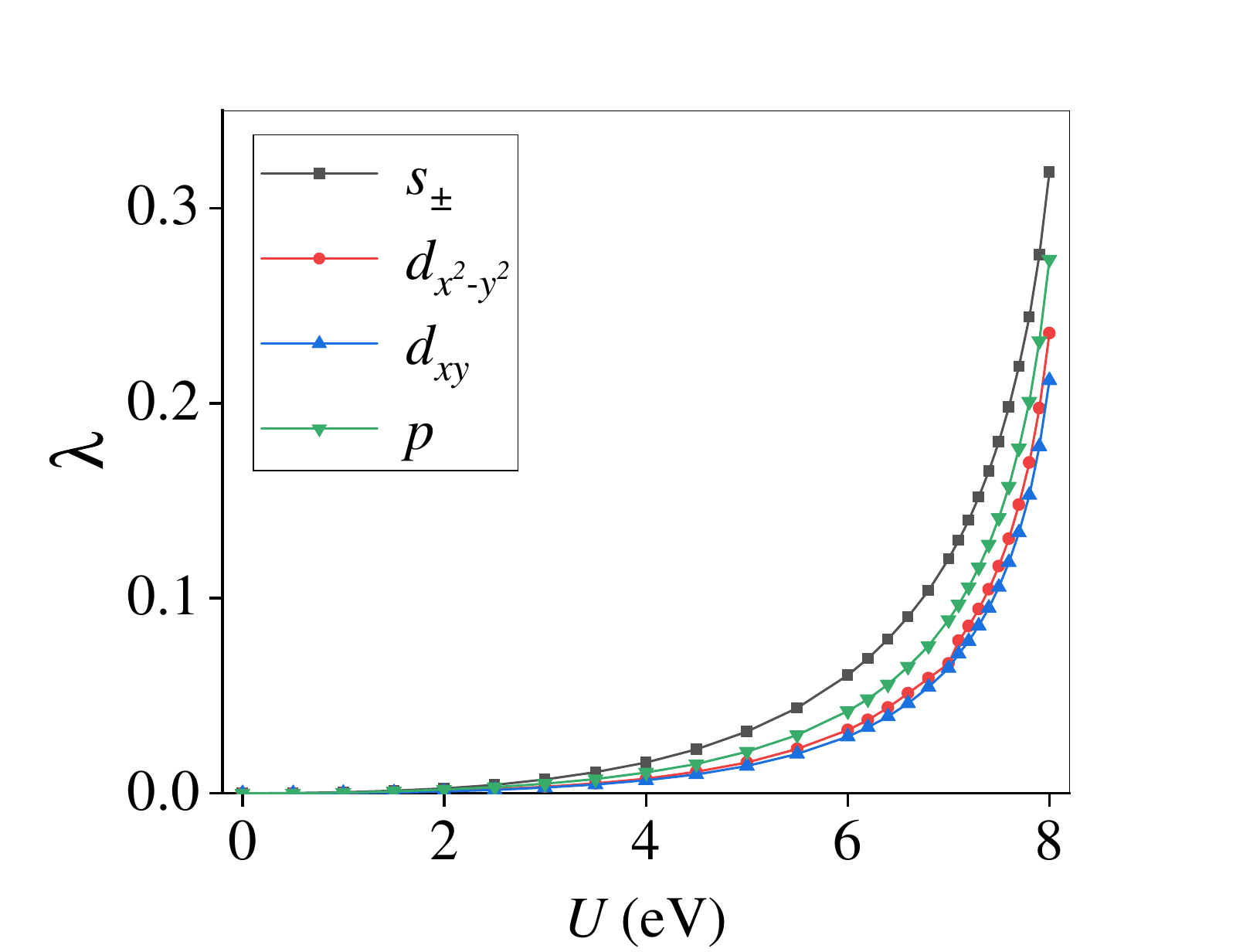}}\subfigure[]{ \label{fig5b}\includegraphics[width=0.23\textwidth]{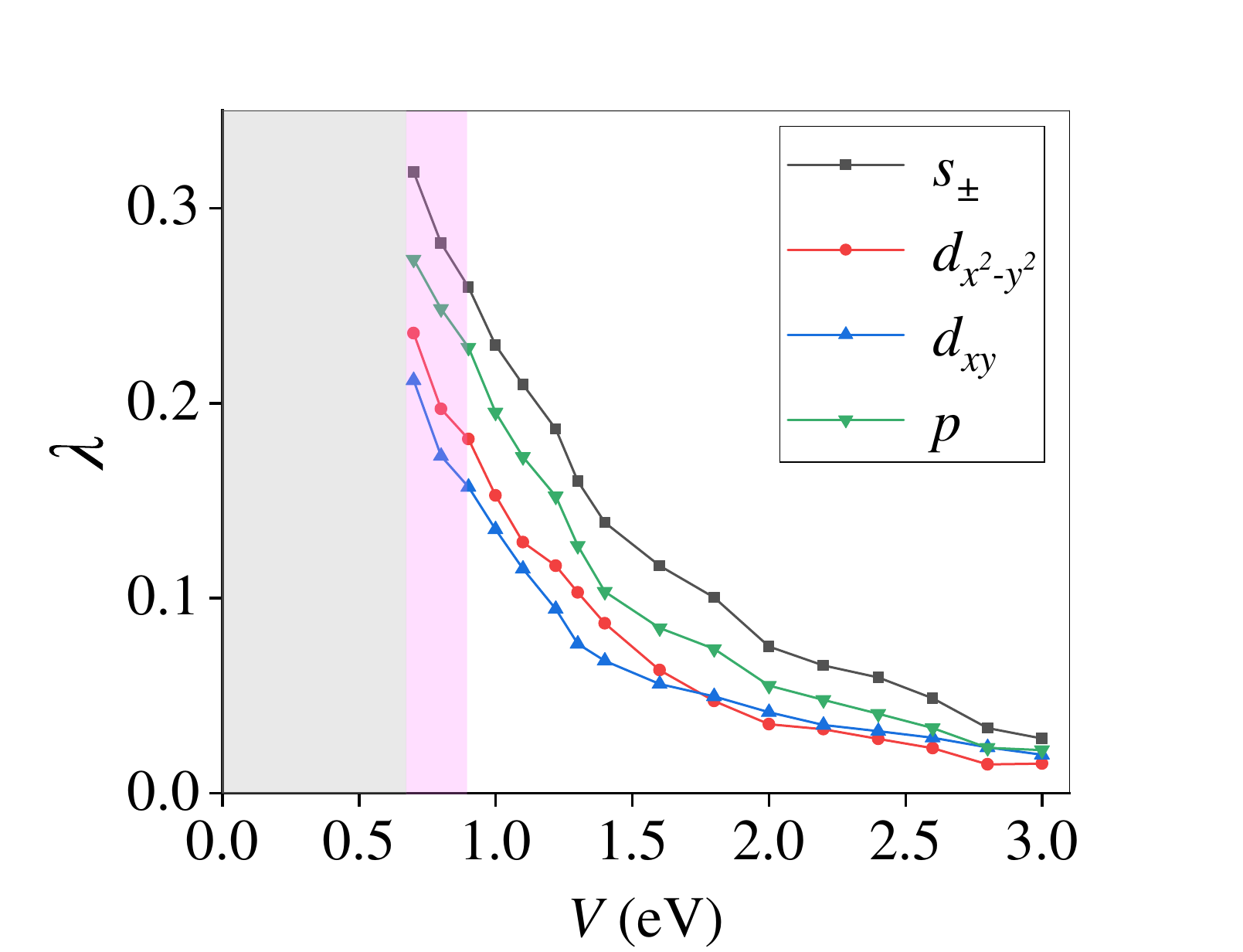}}

  \caption{(a)The largest pairing eigenvalues $\lambda$ in four different pairing symmetry channels as a function of $U$ with $V$ = 0.7 eV. (b)The largest pairing eigenvalues $\lambda$ in four different pairing symmetry channels as a function of $V$ with $U$ = 8 eV.}
\label{fig:lambda}	
  
\end{figure}

To investigate the relationship between interaction strength and pairing symmetry, we examine the $U$-dependence of the largest pairing eigenvalue $\lambda$ for each of the three pairing symmetries. The results are presented in Fig.~\ref{fig:lambda}(a) for $V$ = 0.7 eV. As the interaction strength $U$ increases, $\lambda$ grows, reflecting the enhanced tendency of the system to form a superconducting state. Notably, $\lambda$ exhibits a rapid increase as $U$ approaches a critical value, denoted $U_c$. This behavior is associated with the enhanced AFM spin fluctuations, which are known to strongly enhance pairing interactions in systems near the critical point of an AFM instability. 

The value of $U_c$ can be tuned by adjusting the parameter $V$, as shown in Fig.~\ref{fig3b}, which illustrates how $U_c$ shifts with varying $V$. This tunability is a key feature of the system, as it allows for the exploration of different magnetic and superconducting regimes depending on the strength of the electron-electron interactions.

For graphene-based materials, the interaction parameter $U$ is typically in the order of 10 eV, but its precise value remains a topic of ongoing debate in the literature \cite{castro2009electronic}. In the context of our study, we have chosen a value of $U$ = 8.0 eV for our calculations, which is within the reasonable range for many materials of this class. It is worth noting that, due to the weak-coupling nature of the RPA used in our calculations, the choice of $U$ is crucial in determining the accuracy of the results, especially near the critical point.

Fig.~\ref{fig:lambda}(b) displays the $V$-dependence of the largest eigenvalue $\lambda$ for each pairing symmetry. It is evident that, when $V$ is less than 0.7 eV, the system enters a regime where $U > U_c$, and the RPA approximation is no longer applicable. In this regime, the fluctuations are too strong for the perturbative treatment of RPA to accurately describe the system's behavior. This limitation of the RPA approach highlights the need for more sophisticated methods, such as the inclusion of vertex corrections or non-perturbative techniques, when exploring the superconducting properties of bilayer octagraphene at higher interaction strengths.

\begin{figure}[t]
\centering

\subfigure[]{\label{fig6a}\includegraphics[width=0.23\textwidth]{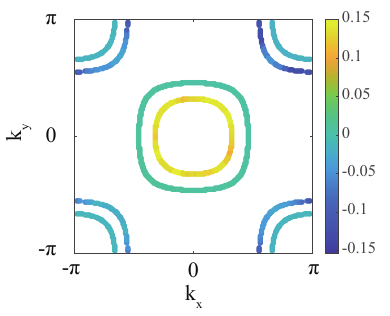}}\subfigure[]{ \label{fig6b}\includegraphics[width=0.23\textwidth]{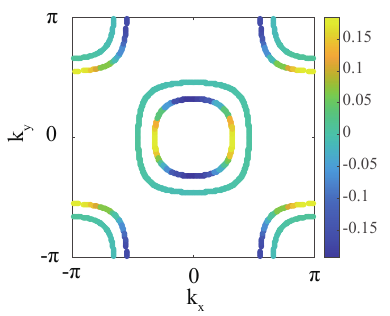}}
	
	\caption{At $V$ = 0.7 eV, the $\textbf{k}$-dependent superconducting order parameter $\Delta_\alpha(\mathbf{k})$ projected onto the FS for (a) $\lambda\approx$ 0.32, which corresponds to $s^{\pm}$-wave pairing and (b) $\lambda\approx$ 0.23, which corresponds to $d_{{x^{2}-y^{2}}}$-wave pairing.}
	\label{fig:paring}
\end{figure}

When the interaction parameter $U$ is set to 8.0 eV, and the value of $V$ is chosen to be 0.7 eV, the maximum eigenvalue $\lambda$ for the superconducting pairing symmetry reaches 0.32. This suggests that bilayer octagraphene, under appropriate conditions, could exhibit superconducting properties, furthering its potential as a candidate material for unconventional superconductivity.

The pairing gap function $\Delta_\alpha(\mathbf{k})$, which describes the superconducting order parameter in momentum space, is determined for the leading superconducting symmetries: the $s^{\pm}$-wave and $d_{x^2 - y^2}$-wave. These two pairing symmetries are shown in the FS plots in Fig.~\ref{fig:paring}, where the $C_{4v}$ symmetry of the bilayer octagraphene system is clearly reflected in the distribution of the gap function. For the $s^{\pm}$-wave symmetry, the gap changes sign between different Fermi pockets. On the other hand, the $d_{x^2 - y^2}$-wave symmetry is associated with a nodal gap structure.

Our calculations indicate that the leading pairing symmetry for bilayer octagraphene under the given conditions is the $s^{\pm}$-wave, as indicated by the largest eigenvalue $\lambda$. The superconductivity in this system is primarily driven by AFM spin fluctuations, similar to what is observed in iron-based superconductors. The $s^{\pm}$-wave pairing symmetry implies a gap function with opposite signs on different Fermi pockets, which is consistent with the behavior expected in systems near an AFM quantum critical point.

However, as the interaction strength $U$ approaches the critical value $U_c$, a significant issue arises. Near this critical point, the system experiences very large spin fluctuations, which arise due to the diverging susceptibility. This phenomenon is illustrated in the purple region of Fig.~\ref{fig:lambda}(b), where $\lambda$ increases rapidly as $U$ nears $U_c$. While this suggests a strong tendency for superconductivity, it also signals a potential overestimation of the pairing eigenvalue $\lambda$, as the RPA used in the calculation becomes less reliable in the presence of such strong fluctuations. In this regime, the perturbative nature of RPA may not accurately capture the true behavior of the system, and more sophisticated approaches, such as the inclusion of vertex corrections or non-perturbative methods, are required to obtain a more accurate description of the superconducting properties.

\section{\label{sec:con}Conclusions}
In this study, we have explored the superconducting properties of bilayer octagraphene under the influence of a perpendicular electric field, with a focus on the interplay between pairing symmetries, electron interactions, and spin fluctuations. We identified that the bilayer octagraphene system, belonging to the $C_{4v}$ point group symmetry, can support both $s^{\pm}$-wave and $d_{x^2 - y^2}$-wave superconducting states, with the $s^{\pm}$-wave pairing emerging as the leading symmetry. Our results indicate that the superconducting pairing eigenvalue $\lambda$ reaches a maximum value of 0.32 for $V$ = 0.7 eV and $U$ = 8.0 eV, suggesting bilayer octagraphene as a promising candidate for unconventional superconductivity.

Superconductivity has been realized in various bilayer graphene systems, including twisted bilayer graphene with moir\'e flat bands\cite{18} and Bernal-stacked bilayer graphene\cite{li2024tunable}. In the latter case, the application of an electric field, particularly in proximity to monolayer \textup{WSe$_2$}, has been shown to induce and modulate superconducting phases in both electron- and hole-doped regimes. Compared with chemical doping, electric field modulation offers a cleaner and more controllable approach, as it does not introduce additional disorder and allows for continuous tuning. Notably, electric-field-induced superconductivity has been experimentally demonstrated even in twisted bilayer graphene systems \cite{park2021tunable,hao2021electric,dutta2024electric}, highlighting the general effectiveness of electrostatic tuning in graphene-based superconductors. In this paper, the interlayer potential energy difference used for calculation is approximately 1 eV, corresponding to an achievable electric field strength of around $10^{9}$V/m\cite{PhysRevLett.99.035001}. The successful synthesis of one-dimensional carbon nanoribbons incorporating four- and eight-membered rings \cite{37}  provides a promising pathway toward the realization of two-dimensional octagraphene. The tunable superconductivity of bilayer octagonal graphene under perpendicular electric field modulation predicted in this paper provides a new idea for the modulation of novel two-dimensional superconducting materials.
Our findings provide valuable insights into the role of electron interactions and the symmetry of the pairing state, which could guide future experimental efforts to synthesize and explore new materials with similar properties.

\begin{acknowledgments}
Y.T.Y. and D.X.Y. are supported by NKRDPC2022YFA1402802, NSFC-12494591, NSFC-92165204, Leading Talent Program of Guangdong Special Projects (201626003), Guangdong Provincial Key Laboratory of Magnetoelectric Physics and Devices (No. 2022B1212010008), Research Center for Magnetoelectric Physics of Guangdong Province (2024B0303390001), and Guangdong Provincial Quantum Science Strategic Initiative (GDZX2401010). F.Y. is supported by the National Natural Science Foundation of China under the Grant Nos. 12234016 and 12074031.
\end{acknowledgments}

\nocite{*}
\bibliography{reference}
\end{document}